\documentclass[pdflatex,sn-mathphys]{sn-jnl}% Math and Physical Sciences Reference Style
\usepackage{amsmath}
\usepackage{gensymb}

\newcommand{\ours}{Twin-S} 
\newcommand{\F}[2]{^{#1}\textrm{F}_{#2}}
\newcommand{\FT}[1]{^{o}\textrm{F}_{#1}}

\newcommand{\db}{db}
\newcommand{\drill}{d}
\newcommand{\pb}{pb}
\newcommand{\phan}{p}

\newcommand{\weightsA}[1]{\beta _{#1}}

\newcommand{\Rot}[2]{^{#1}R_{#2}}

\newcommand{\al}[2]{^{#1}\vec\alpha_{#2}}

\newcommand{\sk}[1]{#1^\wedge}

\makeatletter
\newcommand{\ie}{\emph{i.e.}\@ifnextchar.{\!\@gobble}{}}
\newcommand{\eg}{\emph{e.g.}\@ifnextchar.{\!\@gobble}{}}
\newcommand{\etc}{etc\@ifnextchar.{}{.\@}}
\makeatother

\jyear{2023}%
\UseRawInputEncoding
\theoremstyle{thmstyleone}%
\theoremstyle{thmstyletwo}%

\theoremstyle{thmstylethree}%

\newcommand{\rev}{\color{black}}

\begin{document}

\title[]{\ours{}: A Digital Twin Paradigm for Skull Base Surgery}

%%=============================================================%%
%% Prefix	-> \pfx{Dr}
%% GivenName	-> \fnm{Joergen W.}
%% Particle	-> \spfx{van der} -> surname prefix
%% FamilyName	-> \sur{Ploeg}
%% Suffix	-> \sfx{IV}
%% NatureName	-> \tanm{Poet Laureate} -> Title after name
%% Degrees	-> \dgr{MSc, PhD}
%% \author*[1,2]{\pfx{Dr} \fnm{Joergen W.} \spfx{van der} \sur{Ploeg} \sfx{IV} \tanm{Poet Laureate} 
%%                 \dgr{MSc, PhD}}\email{iauthor@gmail.com}
%%=============================================================%%

\author*[1]{\fnm{Hongchao} \sur{Shu}}\email{hshu4@jhu.edu}
\equalcont{These authors are joint first authors.}

\author[1,2]{\fnm{Ruixing} \sur{Liang}} % \email{rliang7@jh.edu}
\equalcont{These authors are joint first authors.}

\author[1]{\fnm{Zhaoshuo} \sur{Li}} % \email{zli122@jhu.edu}
\equalcont{These authors are joint first authors.}

\author[1]{\fnm{Anna} \sur{Goodridge}}
\author[1]{\fnm{Xiangyu} \sur{Zhang}}
\author[1]{\fnm{Hao} \sur{Ding}}
\author[2]{\fnm{Nimesh} \sur{Nagururu}}
\author[1]{\fnm{Manish} \sur{Sahu}}
\author[2]{\fnm{Francis X.} \sur{Creighton}}
\author[1]{\fnm{Russell H.} \sur{Taylor}}
\author[1]{\fnm{Adnan} \sur{Munawar}}
\author[1]{\fnm{Mathias} \sur{Unberath}}

\affil[1]{\orgname{Johns Hopkins University}, \city{Baltimore}, \state{MD}, \country{United States}}

\affil[2]{\orgname{Johns Hopkins Medicine}, \city{Baltimore}, \state{MD}, \country{United States}}

% \author{Anonymous Submission}

%%==================================%%
%% sample for unstructured abstract %%
%%==================================%%

\abstract{
\textbf{Purpose:} 
Digital twins are virtual replicas of real-world objects and processes. 
This paradigm has great potential in surgical procedures, for example by enhancing the situational awareness of surgeons.
We introduce \textit{\ours{}}, a digital twin framework for skull base surgery.

\textbf{Methods:} 
\ours{} combines high-precision optical tracking and real-time simulation.
To guarantee accurate modeling, \ours{} is enabled by calibration routines to ensure that the virtual representation precisely reflects all real-world processes.
\ours{} models and tracks key elements of skull base surgery, including surgical tools, patient anatomy, and surgical cameras.
Importantly, \ours{} models real-world drilling and updates the virtual model at a frame rate of 28.

\textbf{Results:} 
Our evaluation of \ours{} demonstrates an average error of 1.39 mm during the drilling process. 
Our study also highlights the benefits of \ours{} in downstream applications in image-guided interventions. 
We showcase how \ours{} provides temporally varying augmented overlays  derived from the continuously updated virtual model, thus offering additional situational awareness to the surgeon.

\textbf{Conclusion:} 
Our digital twin paradigm captures the real-world surgical progress and updates the virtual model in real-time through modern tracking technologies. 
Future research integrating vision-based techniques could further increase the accuracy of \ours{}. 
}


\maketitle

\section{Introduction}\label{sec1}
Digital twins are virtual counterparts of real-world processes, modeling dynamics and properties in real-time~\cite{IBM}. 
Receiving continuous measurements from sensor-rich environments, digital twins can conversely provide computational feedback, which is difficult to obtain otherwise. 
Digital twins have been widely adopted in areas of manufacturing, farming, and product design~\cite{jones2020characterising}.
{\rev
In biomedical sciences, digital twins are used in cardiovascular diagnostics~\cite{martinez2019cardio}, insulin pump control~\cite{breton2020randomized}, and \etc.
Digital twins are also envisioned to aid personalized medication~\cite{bjornsson2020digital} and predict human immune system responses~\cite{laubenbacher2022building}. 
}

\begin{figure}[bt]
    \centering
    \includegraphics[width=\textwidth, page=1]{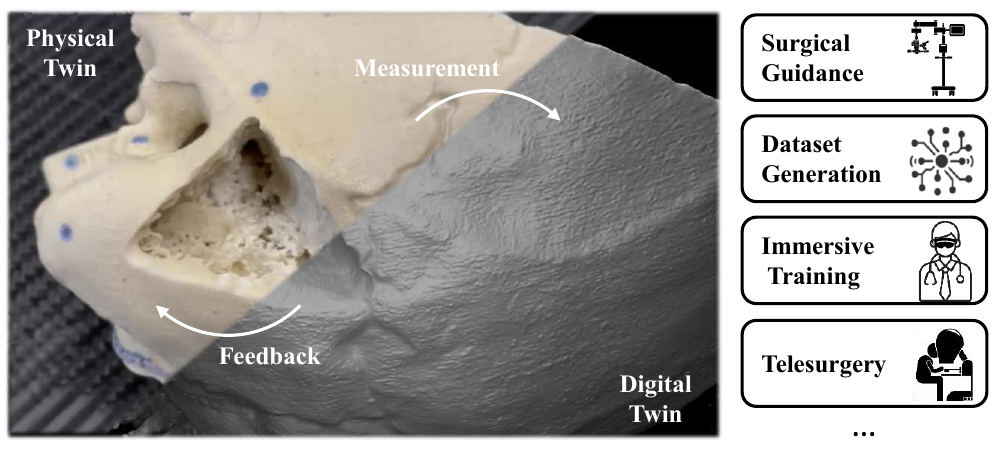}
    \caption{
    In our \ours{} paradigm, measurements are collected from sensors deployed in the operating room and sent to the digital twin continuously. 
    The computation analysis of the simulation feedback to the physical twin.
    The proposed digital twin setup enables many downstream applications across various surgical stages. 
    }
    \label{fig:overview} 
\end{figure}

In surgical scenarios, digital twins can potentially offer advantages across all surgical stages (\autoref{fig:overview}). 
Prior work has explored the use of digital twins in pre-operative planning and surgical training~\cite{coelho2020augmented}. 
When used intra-operatively, digital twins can provide real-time guidance to surgeons for complementary situational awareness and, in turn, facilitate surgical decision-making~\cite{chalasani2016concurrent, wang2017force, yasin2020evaluation}. 
Lastly, digital twins can fully digitize the procedures for record-keeping, post-operative evaluation, and dataset generation for machine learning algorithms.  

{\rev
Some of the earliest works related to the concept of digital twins for surgical assistance date back to 1994. 
RobotDoc~\cite{taylor1994image, kazanzides1999robot}, a robotic system for orthopedic surgeries, visualized CT pre-operative scans and highlighted drilled tissues for guidance.  
Subsequently, many solutions have been presented with more advanced model updating techniques and visualizations. 
For example, Chalasani \textit{et. al.}~\cite{chalasani2018preliminary} and Yasin \textit{et. al.}~\cite{9286681} update the pre-operatively acquired organ model based on force sensing while surgeons palpate the tissue.
Concurrent to our work, Shi \textit{et al.}~\cite{shi_synergistic_2022} uses a digital twin paradigm in liver tumor surgery, where a virtual model is used to predict the motion of the liver for respiratory compensation. 
Digital twin frameworks also have found applications in telesurgery~\cite{laaki_prototyping_2019, bonne_digital_2022}.
}

Augmented Reality (AR) and Virtual Reality (VR) also closely relate to digital twin systems and have been widely adopted for surgical applications~\cite{lungu_review_2021}.
Most prior AR systems overlay the pre-operative scans or plans on the patient anatomy for intra-operative guidance~\cite{aguilar-salinas_current_2022}. 
However, these systems do not update the patient model intra-operatively, limiting their applications~\cite{kockro_dex-ray_2009}.
VR systems, on the other hand, mainly focus on simulations for surgical training~\cite{agha_role_2015}. These VR systems do not receive measurements from the real world, and thus fail to mimic the behaviors of the physical entities in real-time~\cite{munawar_virtual_2021, munawar2023fully}.

We present a digital twin framework named \ours{} for skull base surgeries. 
We specifically consider mastoidectomy~\cite{razavi2019image}, a surgical approach in the lateral skull base whereby bone is removed with surgical drills to obtain access to the middle and inner ear. 
Given the complex arrangement of cortical bone, nerves, vasculature, and end-organs, surgical stereo microscopes are used to navigate in the small operating field.
\ours{} combines high-precision optical tracking and real-time physics-based simulation~\cite{munawar_virtual_2021, munawar2023fully}. 
We rely on calibration routines to ensure that the virtual representation precisely mimics all real-world processes. 
\ours{} models and tracks the critical components of skull base surgery, including the surgical tool, patient anatomy, and surgical camera. 
\ours{} updates the virtual patient anatomy model to account for the real-world tool-to-tissue interactions at a frame rate of 28 FPS.

We conduct experiments to evaluate the accuracy of each component of \ours{}. 
We further derive numerical analysis to provide maximum bounds on tracking error and the contribution to the final error from different stages. 
Finally, we illustrate one use case of \ours{} to improve situational awareness in a mixed-reality setup. 
The contributions can be summarized as follows:
\begin{itemize}
    \item We present a digital twin framework for skull base surgery named \ours{}. It models, tracks and updates all critical components of skull base surgeries in real-time to mimic real-world processes. 

    \item We conduct evaluations on tracking accuracy and simulation fidelity of \ours{}. We present numerical analysis for the worst-case error bounds.
    
    \item We showcase one application of \ours{} in downstream mixed reality tasks to provide surgical guidance and context situational awareness. 
\end{itemize}

\section{System Components}
\label{sec:methods}
\begin{figure}[tb]
    \centering
    \includegraphics[width=\textwidth]{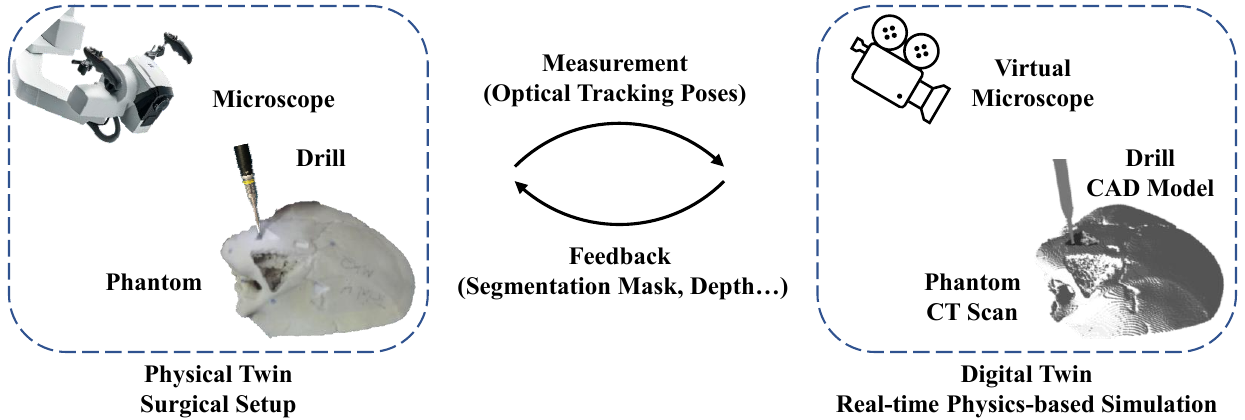}
    \caption{
    \ours{} models, tracks, and updates the surgical drill, phantom, and microscope to emulate the skull base surgical setting.
    Optical tracking provides measurements of object poses to the digital twin.
    Using the CT scan of the phantom and the drill model, a real-time physics-based simulation replicates the real-world processes.
    The simulation generates computational feedback such as segmentation mask and depth back to the physical twin.
    }
    \label{fig:Overview} 
\end{figure}

Building a digital twin system for skull base surgery requires precise modeling, tracking, and updating of the patient's anatomy, the surgeon's tool(s), and the surgical camera.
In our setup, the system includes a surgical phantom to simulate the patient's anatomy, a surgical drill as the ablation tool, and a stereo surgical microscope. 
\ours{} acquires 3D poses of each component via an optical tracker (FusionTrack 500, Atracsys\footnote{\href{https://www.atracsys-measurement.com/products/fusiontrack-500/}{https://www.atracsys-measurement.com/products/fusiontrack-500/}}). 
We use $o$ to denote the optical tracker.
The pose measurements are streamed into a physics-based real-time simulation built upon AMBF~\cite{munawar_virtual_2021, munawar2023fully}. 
The simulation provides computational analysis as feedback to real-world processes. %For example, \ours{} generates per-pixel segmentation information of the surgical scene for downstream tasks. 
%The three poses of interest are the $\sethree$ transformations from the optical tracker to the camera $\FT{\cam}$, the surgical drill $\FT{\drill}$ and the surgical phantom $\FT{\phan}$.
An overview is shown in \autoref{fig:Overview}. 

\subsection{Modeling of the Surgical Drill}
The 3D model of the surgical tool is obtained from its manufacturer (ANSPACH, Johnson \& Johnson\footnote{\href{https://www.jnjmedtech.com/en-EMEA/product/anspach-eg1-electric-system}{https://www.jnjmedtech.com/en-EMEA/product/anspach-eg1-electric-system}}). 
In \ours{}, we define the drill coordinate frame to be the drill tip center. 
We mount the optical tracking markers at the drill shaft's tail, defining the drill's base coordinate.
We need to calibrate the transformation from the \textit{d}rill's \textit{b}ase coordinate to the \textit{d}rill coordinate $\F{\db}{\drill}$.

The translational component $^{\db}\textrm{t}_{\drill}$ is obtained by pivot calibration~\cite{yaniv2015pivot}. 
{\rev
For the rotational component $^{\db}\textrm{R}_{\drill}$, it is only necessary to calibrate two degrees of freedom as the drill is symmetrical along its shaft. 
The calibration reduces to aligning two directions -- the shaft direction in the drill coordinate (denoted as $\Vec{P}$) and in the optical tracker coordinate (denoted as $\vec{Q}$).
Using the Kabsch algorithm~\cite{kabsch1976solution}, the rotation can be recovered as:
\begin{align}
    ^{\db}\textrm{R}_{\drill} = (H^TH)^{\frac{1}{2}}H^{-1}, \ \text{where} \ H = \vec{P}^{T}\vec{Q}.
\end{align}
The shaft direction in the drill coordinate $\Vec{P}$ is defined to be along the Z-axis and thus is known. 
To recover the shaft direction in optical tracker coordinate $\vec{Q}$, we fix the drill on a robot arm end effector, where the drill holder by design aligns the drill shaft with the Z-axis of the robot to sufficient accuracy~\cite{feng2016accuracy}. Therefore, by commanding the robot to move along its Z-axis, we recover the direction of motion $\vec{Q}$ in the optical tracker coordinate. Given known $\vec{P}$ and measured $\vec{Q}$, we can recover $^{\db}\textrm{R}_{\drill}$ up to two degrees of freedom.
}

\subsection{Modeling of the Surgical Phantom}

We obtain the 3D structure of the surgical phantom using a CT scanner (Loop-X, Brainlab\footnote{\href{https://www.brainlab.com/surgery-products/overview-platform-products/robotic-intraoperative-mobile-cbct/}{https://www.brainlab.com/loop-x}}).
The phantom is modeled as a binary volume of occupancy, where voxels corresponding to the bony tissues are marked as occupied and voxels representing air are marked as free space.  
To track the surgical phantom, we rigidly mount the phantom on a polycarbonate board with optical tracking markers, defining the base coordinate of the phantom.
To calibrate the transformation from \textit{p}hantom \textit{b}ase to \textit{p}hantom $\F{\pb}{\phan}$, we directly compute the transformation between the virtual model and the physical phantom via the point-to-plane ICP registration~\cite{schenker1992sensor}. 
We sample 380 points on the physical phantom surface using a tracked pointer tool for the purpose of calibration. 

\subsection{Modeling Tool-to-tissue Interaction}
When surgeons perform drilling in the real world, we update the surgical phantom in real-time. 
We approximate the drill tip as a sphere and detect collisions between the surgical phantom and the drill tip given the tracked positions.
The voxels that collide with the drill tip are moved and set to free space to simulate the tissue removal process. 
More details of the drilling algorithm can be found in prior work~\cite{munawar_virtual_2021}.

\subsection{Modeling of the Surgical Camera}
\label{ssec:camera}
We obtain the intrinsic parameters and distortion coefficients of the stereo camera using a ChArUco pattern. Rectification is then performed to obtain a projective camera model~\cite{zhang2000flexible}. 
To track the surgical camera, optical tracking markers are mounted on the handle of the camera, defining the base coordinate frame. 
A hand-eye calibration routine~\cite{horaud1995hand, furrer2018evaluation} is used to obtain the transformation from the \textit{c}amera \textit{b}ase coordinate to the \textit{c}amera coordinate $^{cb}F{c}$.
Given the tracked camera, \ours{} generates per-pixel segmentation mask and depth maps based on the object information~\cite{munawar_virtual_2021}, which can be used for different downstream applications.

\section{Experiments and Results}
\label{sec:result}
In the following sections, we conduct experiments and numerical analysis on \ours{} to characterize its accuracy.

\subsection{Optical Tracking Accuracy}
\label{ssec:optical_tracking}
\begin{figure}[bt]
    \centering
    \includegraphics[width=\textwidth]{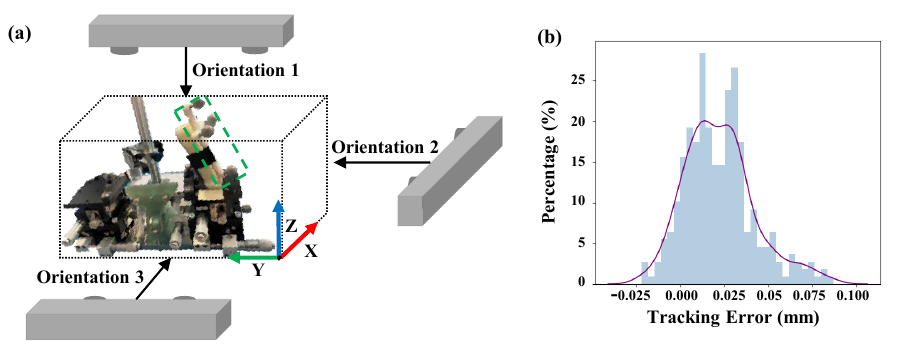}
    \caption{(a) Optical tracker evaluation setup, 4 optical tracking markers are fixed on a 3-axis micrometer stage. Relative orientations of the micrometer stage w.r.t the optical tracker. (c) Histogram of tracking errors.}
    \label{fig:tracker_eval} 
\end{figure}

We conduct experiments to evaluate the accuracy of the optical tracker, as its accuracy directly impacts the results of \ours{}. 
Throughout the experiment, we put the optical tracker at a fixed location. 
We mount the optical markers on a 3-axis micrometer stage (\autoref{fig:tracker_eval} (a)). 
For each experiment trial, we individually move 5 mm along the x, y, and z axis in the micrometer stage local coordinate.
We conduct 3 trials in total, each with a new orientation w.r.t. the optical tracker (\autoref{fig:tracker_eval} (a)). The histogram of error is plotted in \autoref{fig:tracker_eval} (b). 
The mean error is $0.02$~mm, with a standard deviation of $0.02$~mm and a maximum error of $0.08$~mm.
We find the sub-millimeter accuracy of the optical tracker sufficient for our purpose.
We provide additional evaluation results along each axis and the impact of rotation in Appendix A.

\subsection{Camera Calibration Accuracy}
We assess the accuracy of camera calibration as it is critical to downstream applications such as mixed reality and machine learning data generation. 
To perform a quantitative evaluation of hand-eye calibration, we move the camera to different locations observing a static ChArUco pattern.
The image resolution is $1920 \times 1080$.
We take the first video frame as the reference frame. 
The reference frame is then transformed and projected into subsequent video frames.
We evaluate the alignment of the projected pattern and observed pattern and report the re-projection error (RPE) among all frames. 

The hand-eye calibration method only has a $16$~pixels mean RPE, which is equivalent to $0.8\%$ of the image width. 
We further convert the pixel RPE to calibration errors in mm by exploiting the projective relationship between the image space and the Cartesian coordinate using the following equation:
\begin{equation}
    ^{c}\epsilon_u / \text{RPE} \,=\, ^{c}t_u / f_\text{px} .
\end{equation}
{\rev
In other words, the ratio between the calibration error $^{c}\epsilon_u$ in mm and mean RPE in pixels is equal to the ratio between the camera-to-ChArUco distance $^{c}t_u$ in mm and surgical camera focal length $f_\text{px}$ in pixels.}
Given the estimate of $^{c}t_u$ from the physical setup and the calibrated camera focal length $f_{px}$, the hand-eye calibration error converts approximately to $1.9$ mm.

\subsection{Drilling Simulation Accuracy}
\label{ssec:drilling}
\begin{figure}[tb]
    \centering
    \includegraphics[width=\textwidth]{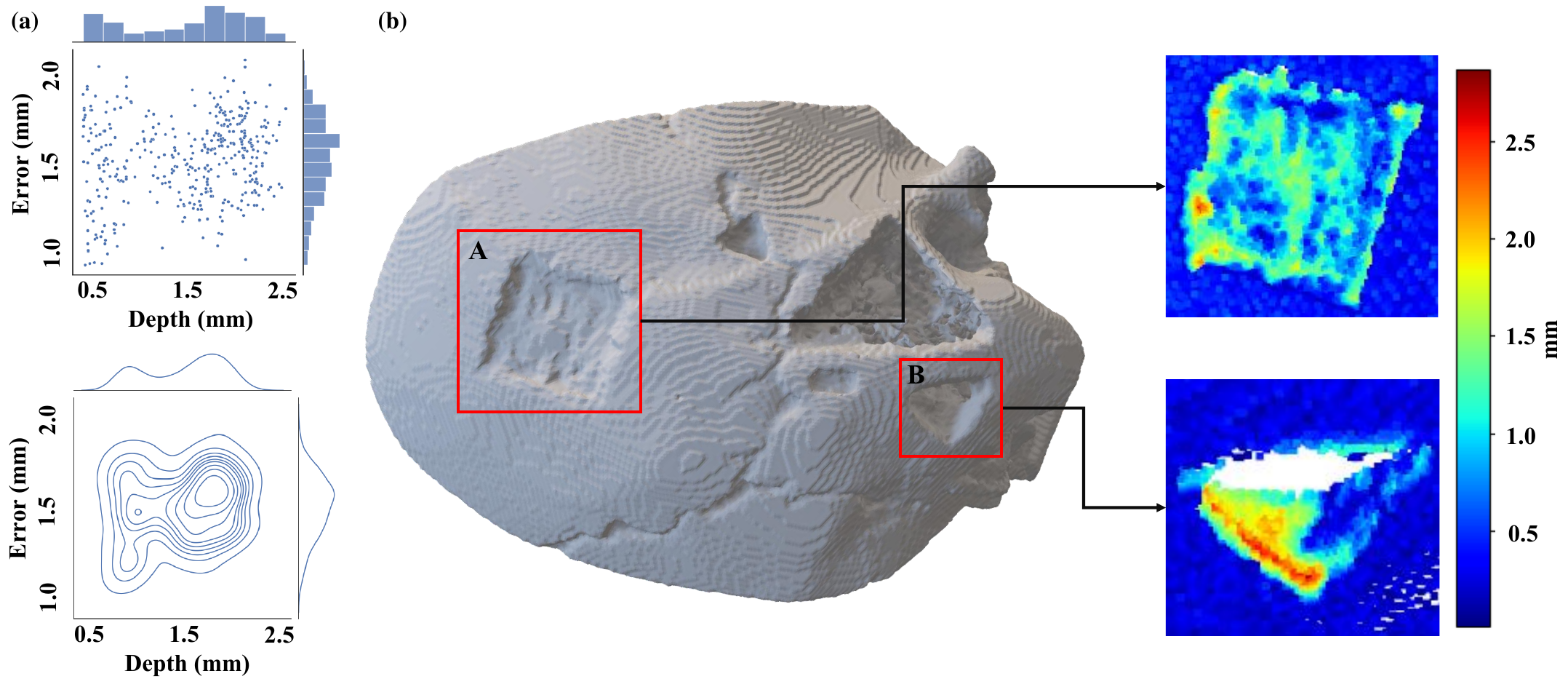}
    \caption{{\rev (a) Scatter and density plots of drilling depth and simulation error. We have not found a statistically significant correlation between drilling depth and error ($p$-value=1). (b) Evaluation of misalignment between ground truth CT scan and a phantom point cloud from the virtual model. The lower error indicates better agreement between the CT scan and the virtually updated model from \ours{}.}}
    \label{fig:depth_error} 
\end{figure}

One of the key features of \ours{} is the real-time updating of the anatomic model as the patient's anatomy is ablated. 
We experiment to evaluate the accuracy of the drilling process. 
A surgical expert is invited to drill the surgical phantom similar to \textit{in vivo}  mastoidectomy. 
We compare the updated phantom model of our Twin-S framework after drilling with the 3D structure obtained by a Loop-X CT scanner. 
We visualize the mismatch between the virtual estimate of the tissue ablation and the true state of the phantom after the surgery as a heat map in \autoref{fig:depth_error} (b). 
{\rev
We report the error in drilled regions only instead of the entire model to evaluate the model update accuracy of \ours{}.
}

The mean error of our virtual model in region A is $1.18$~mm with a standard deviation of $0.24$~mm, and the mean error in region B is $1.61$~mm with the standard deviation of $0.49$~mm. {\rev The average error of all drilled regions is $1.39$~mm with a standard deviation of $0.62$~mm.} Errors are mostly located around sharp edges, which are affected to some extent by the limited spatial resolution of the CT scan.  
{\rev We further examine if there is correlation between the drilling depth and error, with qualitative results shown in \autoref{fig:depth_error} (a).
% Depth is computed from the correspondent coordinates distance between pre-drilling and post-drilling ground truth CT scans via ICP. 
% By exploiting these two correspondences, we could address this correlation 
% where we exclude most points depth around 0 since no drilling is done on those areas. 
We do not observe strong correlations between the two variables qualitatively, and the p-value of 1 from our chi-square test confirms that the variables are indeed statistically independent. 
}

Overall, the results indicate that \ours{} can update the anatomical model with a precision comparable to conventional optical navigation systems~\cite{holland2021hidden}, which is expected since spatial tracking in \ours{} largely relies on those systems. 
{\rev
However, the average drilling depth of 1.57 mm is shallower than clinical practices~\cite{Mastoid}. The line-of-sight issues of the optical tracking system currently prohibit us from drilling deeper. It is our future work to resolve the line-of-sight issue and expand our evaluation in deeper regions.
}

% \begin{figure}[tb]
%     \centering
%     \includegraphics[width=0.7\textwidth]{pointcloud_eval.pdf}
%     \caption{Evaluation of misalignment between ground truth CT scan and a phantom point cloud from the virtual model. The lower error indicates better agreement between the CT scan and the virtually updated model from \ours{}. }
%     \label{fig:pointcloud_eval} 
% \end{figure}

\subsection{Numerically Analyzing Tracking Performance}
We analyze the tracking error to shed insights into sources of error in the tracking performance. 
We denote the measured transformation as $F^{*}$, which contains error $\Delta F$ and deviates from the actual transformation $F$:
\begin{equation}
    F^{*}=F\cdot\Delta F,\
\text{where}\
    \Delta F = 
    \begin{bmatrix}
        I+\sk{\vec\alpha}, & \vec\varepsilon\\ 
        0, & 1
    \end{bmatrix}.
\end{equation}
{\rev
The $\alpha$ is the rotational error using axis-angle notation, and $\sk{(\cdot)}$ is a skew-symmetric matrix given a vector input.
}
The $\vec\varepsilon$ is the translation error.

We present an analysis of the relative pose error between phantom and drill, which is critical for modeling tool-to-tissue interactions. The measured phantom-to-drill transformation is expressed as:
\begin{align}
    \F{\drill}{\phan}^{*} &= \, \FT{\drill}^{*-1} \,\cdot\, \FT{\phan}^*, \nonumber \\
     &=(\FT{\db}^{*}\,\cdot\,\F{\db}{\drill}^{*})^{-1}\cdot \, \FT{\pb}^{*}\,\cdot\,\F{\pb}{\phan}^{*}.
\end{align}
In the following content, we use the rotational component as an example to illustrate our derivation and analysis. We expand the error terms of the rotation part and obtain:
\begin{align}
    \Rot{\drill}{\phan}^* = & \, \Rot{\db}{\drill}^{*-1} \cdot \, \Rot{o}{\db}^{*-1} \cdot \Rot{o}{\pb}^* \cdot \Rot{\pb}{\phan}^*, \nonumber \\
    \Rot{\drill}{\phan}\cdot\big(I+\sk{\al{\drill}{\phan}}\big) = & \bigg(\, \Rot{\db}{\drill} \cdot \big(I+\sk{\al{\db}{\drill}}\big)\bigg)^{-1} \cdot \bigg(\,\Rot{o}{\db} \cdot \big(I+\sk{\al{o}{\db}}\big)\bigg)^{-1} \cdot \nonumber\\
    &\Rot{o}{\pb}\cdot\big(I+\sk{\al{o}{\pb}}\big)\cdot \, \Rot{\pb}{\phan} \cdot\big(I+\sk{\al{\pb}{\phan}}\big).
\end{align}
With re-arrangement, we can obtain 4 independent components contributing to the rotational error:
\begin{equation}
        \vec{^{\drill}\alpha _{\phan}} = 
    \weightsA{1}\,\cdot\,\vec{^{\pb}\alpha _{\phan}}+\weightsA{2},\cdot\,\vec{^{o}\alpha _{\pb}}+\weightsA{3}\,\cdot\,\vec{^{o}\alpha _{\db}}+\weightsA{4}\,\cdot\,\vec{^{\db}\alpha _{\drill}},
    \label{eqn:rot_error}
\end{equation}
where $\weightsA{1}=I$, $\weightsA{2}=\,^{\pb}R_{\phan}^{-1}$, $\weightsA{3}=-\Rot{\pb}{\phan}^{-1}\cdot\,\Rot{o}{\pb}^{-1}\cdot\,\Rot{o}{db}$, and $\weightsA{4}=-\Rot{\drill}{\phan}^{-1}$.
Intuitively, the norm of $\weightsA{i}$ represents how much each measurement error contributes to the final inaccuracy. 

To derive the numerical upper bound on tracking error, we use the worst-case tracking accuracy of 0.08 mm for each optical marker obtained from \autoref{ssec:optical_tracking}. We also assume a worst-case rotation error of $0.5\degree$ for ICP as it is not directly obtainable. We obtain the estimated worst-case error norm as:
\begin{align}
    \left \| \vec{^{\drill}\alpha _{\phan}}  \right \|_2 & \leq \|\weightsA{1}\,\cdot\,\vec{^{\pb}\alpha _{\phan}} \|_2+\|\weightsA{2},\cdot\,\vec{^{o}\alpha _{\pb}}\|_2+\|\weightsA{3}\,\cdot\,\vec{^{o}\alpha _{\db}}\|_2+\|\weightsA{4}\,\cdot\,\vec{^{\db}\alpha _{\drill}}\|_2, \nonumber \\  
    & \leq 0.5+0.3+0.4+0.5=1.7 \degree.
\end{align}
The rotational error $\vec{^{\drill}\alpha _{\phan}}$ is up to $2.7\degree$ and dominated by $\vec{^{\pb}\alpha _{\phan}}$ and $\vec{^{\db}\alpha _{\drill}}$, which are the calibration inaccuracies. A similar analysis can be done on the translation error $\vec{^{\drill}\varepsilon _{\phan}}$, with the worst error bound of 5.29~mm. 

{\rev
The estimated numerical upper bounds of the tracking error represent the worst-case scenario that \ours{} may produce.
In practice, we observe the actual drilling error of 1.39 mm (\autoref{ssec:drilling}) is much smaller than the estimated upper bounds.
}
Details of the translation error analysis are in Appendix B.

\subsection{Segmentation Mask Evaluation}
As mentioned in \autoref{ssec:camera}, \ours{} can generate per-pixel segmentation masks from the virtual models for downstream applications.
Therefore, we evaluate the accuracy of virtually generated segmentation masks and report the Dice scores, where higher Dicer scores indicate better accuracy~{\rev \cite{milletari_v-net_2016}}.
To obtain ground truth, we manually label a sequence of 100 frames. 
The Dice score for the drill is $0.725$, and the Dice score for the phantom is $0.956$. 
The lower Dice score of the drill in comparison to the phantom is due to the thin shape of the drill, where slight offsets will lead to much lower Dice scores.  

\subsection{Runtime Performance Evaluation}
{\rev
We conduct a computation evaluation of Twin-S on a laptop of Ubuntu 20.04 equipped with Nvidia 3060 Laptop Graphic card and Intel i5 CPU.
The breakdown of the computation time is presented in \autoref{table:computation}. The data synchronization step is the most time-consuming process of $20.5$~ms, which aligns the timestamps and pairs the incoming tracking poses from the optical tracker and stereo images from the surgical camera. The overall run time of Twin-S is $35.7$~ms ($28$~FPS).

The calibration step is a requirement prior to launching \ours{} such that the virtual models are aligned with the physical objects. 
Therefore, the calibration process does not impact the runtime performance of Twin-S. 
The time spent for calibration is approximately 10 minutes. 
}

\begin{table}[tb]
\centering
\caption{{\rev Computation evaluation of \ours{}}}
    \begin{tabular}{c|c}
         \textbf{Processes of \ours{}} & \textbf{Mean
         Computational Time (ms)} \\\hline 
        Data synchronization & 20.5\\
        Pose update & 0.2\\
        Simulator volume rendering & 12.0 \\
        Collision simulation & 3.0 \\
        \hline
        Overall & 35.7
    \end{tabular}
    \label{table:computation}
\end{table}

% {\rev Noteworthily, computational time here are the results tested on our platform which is subject to changes when switching to different platforms with other hardware.}

\section{Use Case in Mixed Reality}
\begin{figure}[bt]
    \centering
    \includegraphics[width=\textwidth]{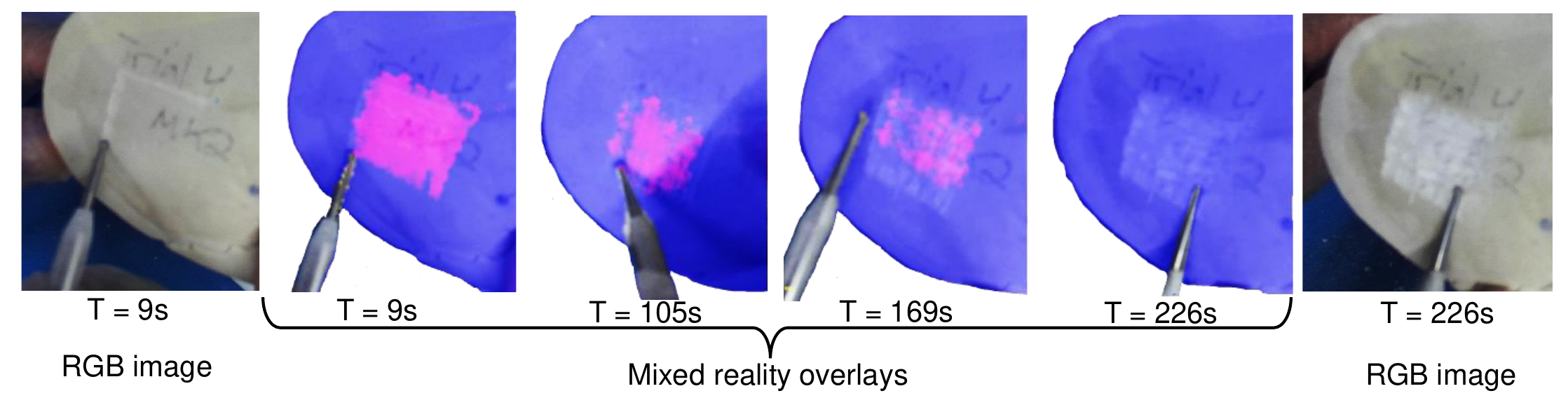}
    \caption{An illustration of using \ours{} for mixed reality intra-operative guidance. The color overlays generated from the virtual model encode the distance to the deep-seated target.}
    \label{fig:MR_demo} 
\end{figure}

We explore a mixed reality use case for providing complementary situational awareness using contextually updated virtual content. 
We consider the scenario where a pre-operative plan is available for the desired bone ablation in the temporal bone. 
We present a temporally adaptive overlay to encode the distance between the current drilled surface and the deeper-seated target. 
In doing so, we offer depth information that may be difficult for surgeons to observe due to bone dust, blood, and \etc. 
To demonstrate this idea with \ours{}, we reverse the drilling process and use the final drilled shape as the ``pre-planned target'' to mimic the targeted application and display the drilling status in previous video frames.
{\rev
Specifically, \ours{} displays a warmer color when the surface is far from the target, and a cooler color as the revealed anatomy is closer to the target. Qualitative visualizations are shown in \autoref{fig:MR_demo}.
Such an experiment conceptually demonstrates how \ours{} can be applied to intra-operative guidance, albeit the simple human-computer interaction paradigm.
It is our future work to build a fully functional mixed-reality system with more effective graphics interfaces.
}

\section{Conclusion}
We present \ours{}, a digital twin framework for skull base surgery. \ours{} models, tracks, and updates virtual counterparts of physical entities in real-time with high accuracy. We present a thorough analysis of the tracking performance and illustrate how \ours{} can be used for downstream applications.

In future work, we plan to integrate vision-based tracking algorithms~\cite{li2022tatoo} to further improve the accuracy of \ours{} rather than relying solely on optical trackers. 
Moreover, \ours{} can generate large-volume of paired data, where microscopic images of the surgical scene are paired with virtually generated labels. 
Through this pairing, we are able to reduce the cost of dataset labeling and avoid the sim-to-real transfer issue commonly faced by synthetic data~\cite{li2021revisiting,li2021sins}. 
It is our future work to demonstrate the application of \ours{} in dataset generation.
Lastly, \ours{} enables model-based control~\cite{li2020anatomical} applications of varying anatomies, thus improving surgical safety and patient outcomes.

\backmatter

% \bmhead{Supplementary information} Supplementary video explaining the proposed method is provided.
% Additional details of numerical analysis are provided.

\bmhead{Acknowledgments} This work was supported in part by Johns Hopkins University internal funds, an agreement between Johns Hopkins University and the Multi-Scale Medical Robotics Centre Ltd., and in part by NIDCD K08 Grant DC019708.

\section*{Declarations}
\bmhead{Conflict of interest} Russell H. Taylor and Johns Hopkins University (JHU) may be entitled to royalty payments related to technology discussed in this paper, and Dr. Taylor has received or may receive some portion of these royalties. Also, Dr. Taylor is a paid consultant to and owns equity in Galen Robotics, Inc.  These arrangements have been reviewed and approved by JHU in accordance with its conflict of interest policy.

\bibliography{sn-bibliography}% common bib file
%% if required, the content of .bbl file can be included here once bbl is generated
% \input main.bbl

%% Default %%
%%\input sn-sample-bib.tex%

\end{document}